\documentclass[12pt]{article}
\newcommand{\bea}   {\begin{eqnarray}}
\newcommand{\eea}   {\end{eqnarray}}
\begin{document}
\renewcommand{\thefootnote}{\fnsymbol{footnote}}

\thispagestyle{empty}
\title{Twist Deformation of \\ Rotationally Invariant Quantum Mechanics}
\author{B. Chakraborty\thanks{{\em e-mail: biswajit@bose.res.in}},
Z. Kuznetsova\thanks{{\em e-mail: zhanna.kuznetsova@ufabc.edu.br}}
~and F. Toppan\thanks{{\em e-mail: toppan@cbpf.br}}
~\\~ \\
{\it $~^{\ast}$ S.N. Bose National Center for Basic Sciences,}\\
{\it JD Block, Sector III, Salt-Lake, Kolkata-700098, India.}\\
{\it $~^{\dagger}$ UFABC, Rua Catequese 242, Bairro Jardim,}\\
{\it  cep 09090-400, Santo Andr\'e (SP), Brazil.}\\
{\it $~^{\ddagger}$ CBPF, Rua Dr.}
{\it Xavier Sigaud 150,}
\\ {\it cep 22290-180, Rio de Janeiro (RJ), Brazil.}
}
\maketitle
\begin{abstract}
Non-commutative Quantum Mechanics in $3D$ is investigated in the framework of the
abelian Drinfeld twist which deforms a given Hopf algebra while preserving its Hopf algebra structure.
\par
Composite operators (of coordinates and momenta) entering the Hamiltonian have to be reinterpreted as primitive elements of a dynamical Lie algebra
which could be either finite (for the harmonic oscillator) or infinite (in the general case).  The deformed brackets of the deformed angular momenta
close the $so(3)$ algebra. On the other hand, undeformed rotationally invariant operators can become, under deformation, anomalous (the anomaly vanishes when the deformation parameter goes to zero). The deformed operators, Taylor-expanded in the deformation parameter, can be selected to minimize the anomaly.
We present the deformations (and their anomalies) of undeformed rotationally-invariant operators corresponding to the harmonic oscillator (quadratic potential), the anharmonic oscillator (quartic potential) and the Coulomb potential.
\end{abstract}
\vfill
\rightline{CBPF-NF-013/09}

\newpage
\section{Introduction}

In a previous work \cite{cct} it was shown that the Wigner's Quantization \cite{wig}, unlike the ordinary quantization based on creation and annihilation operators acting on a Fock vacuum, is compatible with a Hopf algebra structure of its Universal Enveloping (graded)-Lie algebra; it can therefore be regarded as the natural framework to investigate Hopf-algebra preserving, twist-deformations of quantum mechanical systems\footnote{We recall that the Wigner's Quantization
is based on super-Lie algebra valued operators acting on a vacuum state which corresponds to a lowest-weight representation; the ordinary quantization is recovered for a specific value of the lowest weight, which is nothing else than the Wigner's vacuum energy, see \cite{cct} and references therein for details.}. Due to the fact that the ordinary quantization is recovered for a special choice of the Wigner's vacuum energy it is quite important to understand whether
and under which prescription a Hopf algebra structure can be implemented for the ordinary quantization (creation and annihilation operators) as well. This is the viewpoint we are adopting in this paper. Essentially, the first of our results here can be stated as follows: composite operators entering the Hamiltonian and made with Heisenberg algebra operators (coordinates, momenta, the constant $\hbar$) have to be treated as primitive elements (generators) of a dynamical Lie algebra. Their ``composite" nature has to be disregarded and only their commutation relations with respect to the other primitive elements (generators) of the dynamical Lie algebra have to be retained. Within this framework the Universal Enveloping Algebra of the dynamical Lie algebra is endowed with a Hopf algebra structure.
  \par
  The next topic consists in applying an abelian Drinfeld twist which deforms the Universal Enveloping Algebra while preserving its Hopf algebra structure. Since the twist is defined in terms of the three momenta $p_i$, for consistency these generators have to be counted among the primitive elements of the dynamical Lie algebra. A deformed Universal Enveloping Algebra expressed in terms of the twist-deformed primitive elements and their twist-deformed brackets follows from this construction. The next point consists in investigating the behavior of $3D$ non-relativistic quantum mechanical systems which are originally (i.e., in the undeformed case) rotationally invariant. One is guaranteed that the $so(3)$ algebra is preserved by the twisted angular momenta under the twisted brackets. On the other hand those operators which, at the undeformed level, are rotationally invariant (since they commute with the ordinary angular momentum generators under the ordinary brackets) can become anomalous. This means that their twisted commutators with respect to the twisted angular momenta
  can be non-vanishing. The non-zero result, called the deformation ``anomaly", vanishes when the deformation parameter goes to zero. The anomalous operators are expanded in Taylor-series of the deformation parameter ${\vec{\rho}}$. A specific choice of the higher-order contributions can be made in order to minimize the overall anomaly (a similar feature is also encountered for standard quantum anomalies). These considerations apply for both the deformation of (undeformed) rotationally invariant primitive elements, as well as (undeformed) rotationally invariant composite operators (the operator ${\vec L}^2$, which is a Casimir of the $so(3)$ subalgebra, but not a Casimir of the whole Euclidean algebra $e(3)$, is perhaps the most obvious example).
  \par In an Appendix we provide some motivations for the special role played by both the twist-deformed generators and the twist-deformed brackets. On the other hand the connection between the abelian twist deformation and the non-commutative quantum mechanics results from the fact that the ordinary commutator between twist-deformed coordinates gives a constant matrix $\theta_{ij}$. This is a constant element of the Universal Enveloping Algebra and depends on the deformation
  parameter ${\vec \rho}$.
Our work is naturally motivated by the recent upsurge of interest in Noncommutative (NC) theories, both from the condensed matter physics and quantum
gravity point of view. In the former case it has been known for a very long time that the guiding center coordinates of an
electron moving in a plane, but subjected to a constant (i.e. uniform and static) magnetic field, give rise to noncommutativity \cite{dun}.
This can have important consequences for example in QHE \cite{myle}. Besides, it can also arise due to Berry curvature effects
appearing from the breaking of time-reversal symmetry in ferromagnetic systems or from the breaking of spatial-inversion symmetry
in materials like GaAs crystals, as it has been been shown by Xiao et. al. \cite{xiao}.
In both these cases, the noncommutativity is of Moyal type, with time being the ordinary c-numbered variable.
On the other hand, it has been argued by Doplicher et. al. \cite{dfr}, by bringing in considerations of both
general relativity and quantum physics, that the nature of space-time is expected
to be fuzzy at the Planck-length scale. Similar conclusions were also drawn by \cite{sw} from low energy considerations of string theory.
Moyal type of Noncommutativity is one of the simplest types where these features can be realized.
\par
On the other hand, we have still to face the perennial problem of rotational/Lorentz symmetry in NC
theories defined in more than 2D. As already recalled in this paper we shall be basically considering the $3D$
problem, where the basic NC-ty among the spatial coordinates only is given by
\bea
[x_i,x_j]&=&i\theta_{ij}.
\eea
Clearly, the vector dual to $\theta_{ij}$, i.e.
\bea
\theta_i&=&{1\over 2}\epsilon_{ijk}\theta_{jk},
\eea
is pointing towards a particular direction, thereby violating the $SO(3)$ symmetry (note that the $2D$ case is safe from this problem).
Nevertheless,
it has been shown in the literature \cite{wes, abj} that this symmetry can be restored in a Hopf algebraic
setting by using a Drinfeld's twist, such that $\theta_{ij}$ remains invariant under the twisted action
of the rotation. This is in conformity with the usual philosophy of the twisted approach, where the matrix $\Theta= \{\theta_{ij}\}$
is regarded as a (matrix-valued) new constant of Nature like $\hbar, G, c $, etc.  This is in contrast to other approaches followed in
literature (see for example \cite{aar}).
In the relativistic field theory this implies, in a similar manner, that the Poincar\'e
symmetry itself is restored, so that the usual Wigner classification of particles remains unchanged. Since these results
 there has been a flurry of activities in this direction. Despite that,
it was still however not clear how one could investigate the simple QM in this framework\footnote{It should be recalled,
in this context, that the $x_i$'s are operators in QM, while they are mere c-numbered labels for the
continuous degrees of freedom in QFT and are not counted as members of the configuration space
of variables. Consequently, we have to impose the Heisenberg algebra between the coordinates
and the conjugate momenta in QM, i.e. $[x_i,p_j]=i\hbar \delta_{ij}$, but not in the case of
QFT. This distinction carries over to NC-QFT where, although the coordinates are promoted
to the level of operators, they are certainly not valued in the same space as the
field or other composite operators. Moreover, there is no conjugate momentum $p_i$ to the coordinate $x_i$.}.
For instance it was not clear how to define a rotationally invariant potential
even in the above mentioned framework of the twisted Hopf algebra. This is an important question,
considering the fact that the exact solution of the energy-spectrum of a particle, confined in
a noncommutative disc, has already been worked out using the method of piece-wise constant potential \cite{scgv},
which was subsequently used
to study the thermodynamics of a system of particles confined in such a disc \cite{sg}. This analysis had the additional virtue of being
carried on in a purely operatorial level, avoiding the pitfalls associated with the inequivalences between Moyal or Voros star product, which are currently debated in the
literature \cite{glv}.
This gives the main  motivation for the present work, whose main results have been sketched before.\par
The scheme of the paper is as follows. In Section {\bf 2} we link (undeformed) Hopf algebras and Second Quantization, pointing out why some
operators should be regarded as ``primitive elements", while other operators should keep their ``composite" property. In Section {\bf 3} we review the needed facts and formulas concerning the abelian Drinfeld twist. In Section {\bf 4} we discuss the twisted rotations presenting general formulas for
the twisted brackets of the twisted angular momentum. In Section {\bf 5} we present the anomalous twist-deformed commutators for (twisted) primitive elements
such as the quadratic (harmonic), quartic and Coulomb potentials and for the deformation of the ${\vec L}^2$ composite operator. In the Appendix we give heuristic considerations motivating the use of both twisted generators and twisted brackets. Finally, in the Conclusions we make some extra comments on the results here found.

\section{Undeformed Hopf Algebras and Second Quantization}

Before addressing the problem of twisting Hopf algebras in association to NC Quantum Mechanics, we need to learn how to apply undeformed Hopf algebras
to Second Quantization. We work in the framework of the Hopf algebra structure of the Universal Enveloping Algebra of a Lie algebra (the Lie algebra itself is regarded as a dynamical symmetry of a quantum mechanical system). Our discussion has a general validity. For simplicity it will be illustrated with the basic examples of the Euclidean Lie algebras $e(2)$ and $e(3)$.\par
Additive operators (whose eigenvalues in a multi-particle state are the sum of the single-particle eigenvalues) have to be
assumed as ``primitive elements" of the dynamical symmetry algebra (i.e., as generators of the Lie algebra). This is because the additivity of the eigenvalues is encoded
in the undeformed coproduct. Indeed,
\bea
\Delta(\Omega)&=&\Omega\otimes{\bf 1}+{\bf 1}\otimes \Omega
\eea
encodes the additivity of the eigenvalues ($\omega_{1+2}=\omega_1+\omega_2$) for the operator $\Omega$.\\
This remark is still valid if the additive operator under consideration is a Casimir operator.
The Hamiltonian $H$ of a free-particle system is an additive operator. Therefore, $H=\frac{(\vec{p})^2}{2m}$ has to be regarded as a primitive element of the dynamical symmetry algebra despite its ``composite" nature. It is also a Casimir operator of the Euclidean algebra.
From physical considerations we are forced to reject the Hopf algebra equivalence $H={\vec p}^2$ (for simplicity we set $m=\frac{1}{2}$)
which would amount to consider ${\vec p}$ as an element of the Lie algebra, with $H$ beloging to the Enveloping algebra.  This Hopf algebra equality
would imply the unphysical coproduct rule for $H$
\bea\label{wrongcop}
\Delta ( H)&=&\Delta({\vec p}^2)= \Delta({\vec p})\cdot\Delta({\vec p}) = {\vec p}^2\otimes {\bf 1} + {\bf 1}\otimes {\vec p}^2 + 2{\vec p}\otimes {\vec p}\neq H\otimes{\bf 1}+{\bf 1}\otimes H.\nonumber\\
&&
\eea
Note that a relation like (\ref{wrongcop}) makes perfect sense (in physical, as well as in mathematical considerations)
by replacing both ${\vec p}$ with the $3D$ angular momenta ${\vec L}$ and the free Hamiltonian $H$ with the $so(3)$
Casimir operator ${\vec L}^2$. Assuming (as it has to be done) that the components of ${\vec L}$
are Lie-algebra primitive elements, the coproduct
\bea
\Delta ({\vec L}^2)&=& \Delta({\vec L})\cdot\Delta({\vec L}) = {\vec L}^2\otimes {\bf 1} + {\bf 1}\otimes {\vec L}^2 + 2{\vec L}\otimes {\vec L}
\eea
reflects the fact that ${\vec L}^2$ is not an additive operator since, for a composite system, we have that
$({\vec L}_{1+2})^2=({\vec L}_1+{\vec L}_2)^2$.\par
Unlike $H$, which has to be assumed as a primitive element, ${\vec L}^2$ is a genuine composite operator.
As this example shows, the distinction between a ``primitive operator" versus a ``composite operator" cannot be done in purely
mathematical terms. Rather, the mathematical setting has to be accommodated to grasp the physical properties of the system under
investigation.\par
Additive operators have a direct interpretation in terms of their primitive coproducts. Composite operators, such as ${\vec L}^2$,
have no such direct interpretation. In this particular example, the eigenvalues of the composite system are obtained by decomposing
into direct sums the tensor products of the subsystems with the help of the Clebsch-Gordan coefficients.\par
When dealing with the Second Quantization we have to specify at first the single-particle states. This can be done by giving
a complete set
of mutually commuting observables. One should note that these observables can be either ``primitive", as well as ``composite" operators in the sense specifed above.
The discussion can be done in general. It is however useful to work out some specific examples that will be used in the following.
Let us consider the Euclidean Lie algebras $e(2)$ and $e(3)$, respectively.\par
$e(2)$ admits the three generators $p_1,p_2, L$, satisfying the commutation relations
\bea
\relax [p_1,L]&=& -ip_2,\nonumber\\
\relax [p_2,L]&=&ip_1,\nonumber\\
\relax [p_1,p_2]&=&0.
\eea
$e(2)$ admits only one Casimir operator, ${\cal C}\equiv{\vec p}^2={p_1}^2+{p_2}^2$. Indeed $\relax[{\vec p}^2,L]=0$. \par
The Casimir corresponds to the energy $E$ of a non-relativistic, free, two-dimensional particle (whose mass has been normalized, as before, to $m=\frac{1}{2}$).
Since the free energy is an additive operator, the Casimir operator ${\cal C}$ has to be added to the dynamical symmetry Lie algebra. For that we need to enlarge $e(2)$
by defining
\bea
{\overline{e(2)}}&=& e(2)\oplus u(1),
\eea
whose primitive generators are $\{p_1,p_2,L,{\cal C}\}$. ${\cal C}$ can be consistently identified with ${\vec p}^2$ as far as Lie-algebra and single-particle eigenvalues are concerned. We force this identification by imposing that the set of mutually commuting operators $p_1, p_2, {\cal C}$ admits the compatible set of respective eigenvalues $(\sqrt{E}\cos\alpha, \sqrt{E}\sin\alpha, E)$. This is not enough to completely specify the states in the Hilbert space because we still need to take into account the information carried on by the angular momentum $L$ (whose eigenvalues are the integers $m$). This can be done as follows. At first, without loss of generality, we fix the ``reference frame" specified by the eigenvalues $p_1=0$, $p_2=\sqrt{E}$
(recovered by setting $\alpha=\frac{\pi}{2}$). Next, we consider the little (Lie) group of transformations respecting the reference frame and their associated Lie-algebra, Hermitian, operators. We can now find a complete set of observable operators which are mutually ``weakly commuting" when the
reference frame constraint is taken into account. In the example above, mutually ``weakly commuting" observables are given by $p_2, L$, since
$\relax [p_2,L]=ip_1\approx 0$, when $p_1\equiv 0$ is taken into account.
\par
We can extend these considerations to the less trivial case of the three-dimensional Euclidean algebra $e(3)$, whose generators ($p_1,p_2,p_3,L_1,L_2,L_3$) satisfy the commutation relations
\bea\label{eucl3}
\relax [p_i,p_j]&=&0,\nonumber\\
\relax [p_i,L_j]&=& i\epsilon_{ijk}p_k,\nonumber\\
\relax [L_i,L_j]&=&i\epsilon_{ijk}L_k
\eea
(the $L_i$'s are the generators of the $so(3)$ subalgebra). \par
$e(3)$ admits two Casimir operators, ${\cal C}_1$, ${\cal C}_2$, given respectively by
\bea\label{twocasimir}
{\cal C}_1&=& {\vec p}^2,\nonumber\\
{\cal C}_2 &=& {\vec L}{\vec p}.
\eea
One should note that ${\vec L}^2$ is a Casimir operator of the $so(3)$ subalgebra; on the other hand it is not a Casimir operator for $e(3)$.\par
We can repeat the same construction as in the $e(2)$ case, enlarging the algebra to ${\overline{ e(3)}}$, by the addition of ${\cal C}_1$, ${\cal C}_2$ as primitive elements,
\bea\label{exte3}
{\overline {e(3)}}&=&e(3)\oplus u(1)\oplus u(1).
\eea
The identification (\ref{twocasimir}) is assumed to hold in the Lie algebra sense, but not in the Hopf algebra sense.\par
By setting $E$ (the energy) to be the eigenvalue of the ${\cal C}_1$ Casimir operator, without loss of generality we can work within the
$p_1=0, p_2=0, p_3=\sqrt{E}$ reference frame. In this reference frame ${\vec L}{\vec p}\equiv L_3p_3$, such that its eigenvalues are expressed by
$m\sqrt{E}$. A set of mutually ``weakly commuting" observables, respecting the given reference frame, is given by
$p_3, L_3, {\vec L}^2$, with eigenvalues $\sqrt{E}$,$ m$, $l(l+1)$, respectively. Indeed
\bea
\relax [p_3, L_3]&=&0,\nonumber\\
\relax[ L_3, {\vec L}^2]&=& 0,\nonumber\\
\relax [p_3, {\vec L}^2]&\approx&0.
\eea
A state of the system is uniquely specified in terms of its free energy $E$, the orbital angular momentum $l$ and its component along the third axis $m$.\par
In the set of three, mutually weakly-commuting operators which specify the state of the (single-particle) system, two of them ($p_3, L_3$) are primitive operators, while the remaining one (${\vec L}^2$) is a composite operator.

\section{The abelian Drinfeld twist}

In this Section we will recall the basic formulas concerning the abelian Drinfeld twist deformation of
the Universal Enveloping Algebra ${\cal U}({\bf g})$ of a given Lie algebra ${\bf g}$. For our purposes the twist is expressed by
${\cal F}\in {\cal U}({\bf g})\otimes {\cal U}({\bf g})$, such as
\bea\label{abeliantwist}
{\cal F}&=& \exp (i\rho_{ij} p_i\otimes p_j),\nonumber\\
\rho_{ij}&=& \epsilon_{ijk}\rho_k,
\eea
where ${\vec \rho}$ is a dimensional $c$-number and $p_i$ ($i,j=1,2,3$) are the three-dimensional momenta.
It is obviously required that $p_i\in {\bf g}$.\par
The twist induces a deformation in the Hopf algebra  ${\cal U}({\bf g})\rightarrow {\cal U}^{\cal F}({\bf g})$ (see \cite{asc}). Particularly, the co-structures are deformed. The deformed co-structures (coproduct, counit and antipode), applied to an element $g\in {\bf g}$, are respectively given by
\bea
\Delta^{\cal F} (g)&=& {\cal F} \Delta (g) {\cal F}^{-1},\nonumber\\
\varepsilon^{\cal F} (g) &=& \varepsilon (g),\nonumber \\
S^{\cal F} (g) &=& \chi S(g) \chi^{-1},
\eea
where
\bea
\chi &=& f^\alpha S(f_\alpha)\in {\cal U}({\bf g})
\eea
(we are denoting, as usual, ${\cal F}=f^\alpha\otimes f_\alpha$, ${\cal F}^{-1} = {\overline f}^\alpha\otimes {\overline f}_\alpha$).\par
The generators of ${\cal U}^{\cal F}({\bf g})$ are expressed as
\bea\label{defgen}
g^{\cal F}&=& {\overline f}^\alpha (g){\overline f}_\alpha.
\eea
The ${\cal F}$-deformed brackets in ${\cal U}^{\cal F}({\bf g})$ are defined through
\bea\label{Fbrackets}
\relax [g^{\cal F}, h^{\cal F}]_{\cal F} &=& {g^{\cal F}}_1 h^{\cal F}S(g^{\cal F})_2,
\eea
where the Sweedler's notation
\bea\label{sweed}
\Delta^{\cal F}(g^{\cal F}) &=& (g^{\cal F})_1\otimes ({g}^{\cal F})_2
\eea
 has been used.\par
 The ${\cal F}$-deformed brackets satisfy the Jacobi identity.\par
 A more complete list of the properties of the twist-deformed Hopf algebra ${\cal U}^{\cal F}({\bf g})$ is encountered in \cite{asc}.
\par
The Universal Enveloping Algebra of the following Lie algebras can be deformed in terms of the (\ref{abeliantwist}) abelian twist. We have\par
{\em i}) the Heisenberg algebra $h_B(3)$, whose generators are $\hbar, x_i, p_i$ (for $i=1,2,3$). $\hbar$
is a central element and the only non-vanishing commutation relations are given by
\bea\label{heiscomm}
\relax [x_i,p_j]&=& i\delta_{ij}\hbar;
\eea

{\em ii}) the Euclidean algebra $e(3)$ (considered in the previous Section), whose generators are $p_i, L_i$.
Its non-vanishing commutators are given by
\bea
\relax [p_i, L_j]&=& i\epsilon_{ijk} p_k,\nonumber\\
\relax [L_i,L_j]&=& i\epsilon_{ijk}L_k.
\eea
This algebra can be induced by the $h_B(3)$ Heisenberg algebra after setting
\bea\label{angmom}
L_i &=& \frac{1}{\hbar}\epsilon_{ijk}x_jp_k
\eea
and interpreting the $L_i$'s as primitive elements\footnote{As recalled in the previous Section, the notion of ``primitive elements'' is used to underline the fact that the generators of the Lie algebra should not be regarded as composite operators of the Hopf algebra structure.}. Similarly, the extended algebra ${\overline {e(3)}}$ introduced in (\ref{exte3}) can be twist-deformed under (\ref{abeliantwist});
\par
{\em iii}) the algebra $g$, whose primitive elements are the Heisenberg algebra generators ${\hbar, x_i, p_i}$ and the angular momentum generators $L_i$ whose commutation relations, as before, can be induced by the
(\ref{angmom}) position;
\par
{\em iv}) the ``oscillator'' algebra ${osc}$, given by the set of primitive elements \\
$\hbar, x_i, p_i, L_i,
H,K,D$. The commutators involving the generators $H,K,D$ can be read from the positions
\bea
H&=& \frac{1}{\hbar} {\vec p}^2,\nonumber\\
D&=& \frac{1}{2\hbar}({\vec x}{\vec p}+{\vec p}{\vec x}),\nonumber\\
K&=& \frac{1}{\hbar}{\vec x}^2
\eea
(the $\hbar$ at the denominator in the r.h.s. expressions is required in order to compensate the corresponding term coming from the (\ref{heiscomm}) commutators).
$H,K,D$ defines the $sl(2)$ subalgebra. The complete set of non-vanishing commutators among primitive elements of $osc$ is given by
\bea\label{ocomm}
\relax [x_i,p_j]&=&i\delta_{ij}\hbar,\nonumber\\
\relax [x_i, L_j] &=&i\epsilon_{ijk}x_k,\nonumber\\
\relax [p_i, L_j]&=& i\epsilon_{ijk}p_k,\nonumber\\
\relax [x_i, H] &=& 2ip_i,\nonumber\\
\relax [x_i,D]&=& ix_i,\nonumber\\
\relax [p_i, D] &=& -i p_i,\nonumber\\
\relax [p_i,K]&=& - 2 i x_i,\nonumber\\
\relax [H, D]&=& -2i H,\nonumber\\
\relax [H,K]&=& -4 i D,\nonumber\\
\relax [D,K]&=& - 2i K.
\eea
The Hamiltonian of the harmonic oscillator is given by a linear combination of $H$ and $K$;

{\em v}) a finite Lie algebra $g_b$ of hermitian operators which can all be regarded as primitive elements and recovered from at most bilinear combinations in ${\vec x}$ and ${\vec p}$, is given by
the set of generators $\hbar, x_i, p_i, P_{ij}, X_{ij}, {M^+}_{ij}, {M^-}_{ij}$. The commutation relations
involving $P_{ij}, X_{ij}, {M^+}_{ij}, {M^-}_{ij}$ can be read by assuming
\bea
P_{ij} &=& \frac{1}{\hbar}p_ip_j,\nonumber\\
X_{ij} &=& \frac{1}{\hbar}x_ix_j,\nonumber\\
{M^+}_{ij} &=&\frac{1}{\hbar}( x_ip_j+p_jx_i),\nonumber\\
{M^-}_{ij}&=& \frac{i}{\hbar}(x_ip_j-p_jx_i);
\eea

{\em vi}) the above construction can be further generalized. Any Lie algebra containing $h_B(3)$ as a subalgebra and at least one primitive element which is expressed as a trilinear (or $k$-linear, for $k\geq 3$) combination in ${\vec x}$ and ${\vec p}$ is necessarily infinite-dimensional. Indeed, the closure of the commutation relations of this generator with the previous ones requires that new higher-order multilinear terms have to be included as primitive elements. This procedure never stops, leading to an infinite-dimensional Lie algebra. This algebra can be regarded as the unfolded algebra
of primitive elements (the multilinear combinations in terms of ${\vec x}$ and ${\vec p}$ is its folded version, in analogy of what happens, in a different context, with finite $W$-algebras \cite{dBT} or the unfolded version of higher-spin algebras,
see \cite{vas}). One should note that a primitive element which is $k$-linear in ${\vec x}, {\vec p}$, requires the $\frac{1}{\hbar^{k-1}}$ factor (for instance, a primitive element can be associated to $\frac{1}{\hbar^3}({\vec x^2})^2$).

\section{Twisted rotations}

The abelian Drinfeld twist (\ref{abeliantwist}) induces, through eq. (\ref{defgen}), the following deformation of the space coordinates
\bea\label{xf}
{x_i}^{\cal F} &=& x_i-\epsilon_{ijk}\rho_k \hbar p_j.
\eea
This deformation corresponds to the Bopp shift and one should note that the second term in the r.h.s. is quadratic in the Heisenberg algebra generators.
This result was also obtained in \cite{cct}.
The shift maps $x_i\in h_B(3)$ into ${x_i}^{\cal F}\in {\cal U} (h_B(3))$.
Concerning the $p_i$ momenta, they undergo no deformation: $p_i^{ \cal F}=p_i$.
\par
The non-commutative quantum mechanics (for a constant operator $\theta_{ij}$) is recovered from the abelian twist. Indeed
\bea\label{ncemer}
[{x_i}^{\cal F},{x_j}^{\cal F}] &=& i\theta_{ij},
\eea
where
\bea\label{nctwist}
\theta_{ij}&=& 2\hbar^2\epsilon_{ijk}\rho_k,
\label{theta-rho}
\eea
with $\theta_{ij}$ an operator belonging to ${\cal U}(h_B(3))$.

Similarly, but in the ``opposite'' direction, the ${\cal F}$-commutator of the ordinary coordinates produces
\bea
[x_i,x_j]_{\cal F} &=& -\frac{1}{2}i\theta_{ij}.
\eea

The ${\cal F}$-commutator among twisted space coordinates is vanishing
\bea
\relax [{x_i}^{\cal F},{x_j}^{\cal F}]_{\cal F} &=& 0.
\eea
The twisted coproduct of the space coordinates and of the twisted space coordinates is respectively given by
\bea
\Delta^{\cal F} (x_i) &=& x_i\otimes {\bf 1}+{\bf 1}\otimes x_i +\epsilon_{ijk}\rho_k(\hbar\otimes p_j-p_j\otimes \hbar), \nonumber\\
\Delta^{\cal F} ({x_i}^{\cal F}) &=& {x_i}^{\cal F}\otimes {\bf 1}+{\bf 1}\otimes {x_i}^{\cal F} -2\epsilon_{ijk}\rho_k p_j\otimes \hbar.
\eea

If the algebra admits as primitive elements, besides the $p_i$'s, the angular momentum operators $L_i$, their deformation ${L_i}^{\cal F}$, induced by the (\ref{abeliantwist}) twist is given by
\bea\label{defL}
{L_i}^{\cal F} &=& L_i+K_i,\nonumber\\
K_i&=& \rho_kp_ip_k-\rho_ip_kp_k.
\eea
The extra-term $K_i$ can also be written as
\bea
K_i&=& -\rho_j{\vec p}^2\Pi_{ij},
\eea
in terms of the $\Pi_{ij}$ projector
\bea
\Pi_{ij}&=& (\delta_{ij}-\frac{p_ip_j}{{\vec p}^2}).
\eea
The twisted coproduct of the (twisted) angular momentum reads as
\bea
\Delta^{\cal F} (L_i) &=& L_i\otimes {\bf 1}+{\bf 1}\otimes L_i +\rho_{k}(p_i\otimes p_k-p_k\otimes p_i),
\nonumber\\
\Delta^{\cal F} ({L_i}^{\cal F}) &=& {L_i}^{\cal F}\otimes {\bf 1}+{\bf 1}\otimes {L_i}^{\cal F} +2\rho_{k}p_i\otimes p_k-2\rho_i p_k\otimes p_k.
\eea

The original $su(2)$ rotational algebra is recovered in terms of the ${\cal F}$-commutator of the twisted angular momentum. We have indeed
\bea
[{L_i}^{\cal F},{L_j}^{\cal F}]_{\cal F} &=& i\epsilon_{ijk}{L_k}^{\cal F}.
\eea
As a consequence we get the first result, namely that the rotational symmetry is preserved by the (\ref{abeliantwist}) twist-deformation.\par
One can also check that
\bea
\relax [x_i^{\cal F},L_j^{\cal F}]_{\cal F}&=&i\epsilon_{ijk}x_k^{\cal F}, \nonumber\\
\relax [p_i^{\cal F},L_j^{\cal F}]_{\cal F}&=&[p_i,L_j^{\cal F}]_{\cal F}=
i\epsilon_{ijk}p_k^{\cal F},
\eea
showing that both $x_i^{\cal F}$ and $p_i$ have vectorial transformation properties under the deformed brackets. The situation is thus different
from the case of the undeformed brackets where $x_i^{\cal F}$, unlike $p_i$, fails to transform as a vector \cite{cct}.\par
 The next important point is to check whether the operators which are rotationally invariant in the undeformed case, keep the rotational invariant property even in the deformed case or otherwise acquire an anomalous term which disappears in the limit ${\vec \rho}\rightarrow 0$.
We investigate, specifically, the commutation relations
\bea\label{defcomm}
&\relax [ {L_i}^{\cal F}, B^\sharp]_{\cal F}&
\eea
for an operator $B^\sharp$ belonging to the Universal Enveloping Algebra of a Lie algebra containing the Euclidean algebra $e(3)$ as a subalgebra and such that $B^\sharp$ is expanded in ${\vec \rho}$ Taylor series:
\bea\label{taylor}\label{taylorexp}
B^\sharp&=& B_0+B_1+B_2+\ldots,
\eea
with $B_k$ $k$-linear in ${\vec \rho}$. Here $B_0\equiv B$ denotes the undeformed limit for ${\vec \rho}\rightarrow 0$ of $B^\sharp$
(we can therefore say that the operator $B^\sharp$ is the deformation of $B$).\par
The rotational invariance in the undeformed limit requires that the following relation involving ordinary commutators and angular momentum operators has to be satisfied
\bea
\relax [L_i, B_0] &=& 0.
\eea

With a little algebra one can easily prove that the deformed commutator (\ref{defcomm}) can be expressed in terms of
ordinary commutators:
\bea\label{defcommexp}
\relax [ {L_i}^{\cal F}, B^\sharp]_{\cal F}&=& [L_i-K_i,B^\sharp] + M_{ik}[p_k,B^\sharp],
\eea
where $K_i$ enters (\ref{defL}) and $M_{ik}$ is given by
\bea
M_{ik} &=& 2\rho_kp_i-2\rho_ip_k.
\eea
The r.h.s. in (\ref{defcommexp}) is a consequence of the equality
\bea
2K_i-M_{ik}p_k&=&0.
\eea
The meaning of the ${\cal F}$-deformed brackets for non-commutative theories is discussed in the Appendix.\par
It is worth pointing out that the Taylor-expanded series (\ref{taylor}) starting with $B_0$ does not necessarily coincide with the ${\cal F}$-deformed operator ${B_0}^{\cal F}$ (which can also be understood as
Taylor-expanded). In the next Section we will discuss this point in more detail.\par
For completeness we write here the ${\cal F}$-deformed operators $H^{\cal F}, D^{\cal F}, K^{\cal F}$ obtained by applying the
(\ref{abeliantwist}) twist to the $H,D,K$ primitive elements of the oscillator algebra $osc$ given in (\ref{ocomm}).
We have
\bea
H^{\cal F} &=& H,\nonumber\\
D^{\cal F} &=& D, \nonumber\\
\relax K^{\cal F} &=& K-2\epsilon_{ijk}\rho_kx_ip_j+\hbar [{\vec \rho}^2{\vec p}^2 -({\vec \rho}{\vec p})^2].
\eea

\section{Anomalous operators}

For our purposes it is useful to set
\bea
\relax L_i (B_n) &=& [L_i, B_n],\nonumber\\
\relax T_i(B_n) &=& -[K_i,B_n]+M_{ik}[p_k,B_n].
\eea
An undeformed rotationally invariant operator $B$ such that
\bea
\relax [{L_i} , B] &=& 0
\eea
can develop, under deformation, an anomaly $A_i$ which is expressed through
\bea\label{anomaly}
\relax [{L_i}^{\cal F} , B^\sharp]_{\cal F} &=& A_i
\eea
(as discussed in the previous Section, $B^\sharp$ is the deformation of $B$).\par
The anomaly $A_i$ can be expanded in powers of the deformation parameter ${\vec \rho}$. We have
\bea\label{iteranom}
\relax (A_i)_0&=& L_i(B_0)=0,\nonumber\\
\relax (A_i)_n &=& L_i(B_n) + T_i(B_{n-1}),
\eea
with $(A_i)_n$ the $n$-th order contribution in ${\vec \rho}$.\par
Let us consider now the deformation of the rotationally invariant primitive elements $H,D,K$ of
the ${{osc}}$ oscillator algebra (\ref{ocomm}). We get that
\bea
&H^\sharp=H^{\cal F}=H&
\eea
is rotationally invariant under twist-deformed rotations since
\bea
\relax [{L_i}^{\cal F} , H^{\cal F}]_{\cal F} &=& 0.
\eea
Similarly,
\bea
&D^\sharp=D^{\cal F}=D&
\eea
is rotationally invariant under twist-deformed rotations since
\bea
\relax [{L_i}^{\cal F} , D^{\cal F}]_{\cal F} &=& 0.
\eea
On the other hand we get that $K$ gets anomalous since
\bea
K^\sharp&=&K-2\epsilon_{ijk}\rho_kx_ip_j-\hbar ({\vec \rho}{\vec p})^2+\gamma\hbar
{\vec \rho}^2{\vec p}^2,
\eea
which coincides with $K^{\cal F}$ for the special value $\gamma=1$, namely
\bea
{K^\sharp}|_{\gamma=1}&=&K^{\cal F},
\eea
is such that
\bea\label{constanom}
\relax [{L_i}^{\cal F} , K^{\sharp}]_{\cal F} &=& 4\hbar \rho_i.
\eea
The r.h.s. term $4\hbar\rho_i$, which is independent of $\gamma$, is the (constant) anomalous operator.\footnote{In this context
it should be recalled that $\rho_i$,
despite its appearance,
transforms as a scalar under rotations. In the same spirit, despite its appearance, $({\vec \rho} {\vec p})$
is not a scalar under
rotations.}
\par
The analysis of the anomaly can be performed also for composite operators. Let us consider the
Euclidean algebra $e(3)$ defined in (\ref{eucl3}). We investigate the ${({\vec L}^2)}^\sharp$ deformation of the composite operator
${\vec L}^2$, which is the Casimir of the $so(3)$ subalgebra. In accordance with the (\ref{taylorexp}) expansion and the (\ref{defcommexp})
equation, the ${({\vec L}^2)}^\sharp$ Taylor-expansion in ${\vec\rho}$ stops at the second order for a minimal anomaly. We indeed get
\bea
({\vec L}^2)^\sharp &=& ({\vec L}^2)_0 +({\vec L}^2)_1 +({\vec L}^2)_2,
\eea
with
\bea
({\vec L}^2)_0&=& {\vec L}^2,\nonumber\\
({\vec L}^2)_1&=& \alpha_1 ({\vec \rho}{\vec p})({\vec p}{\vec L}) +\alpha_2 {\vec p}^2({\vec \rho}{\vec L}),\nonumber\\
({\vec L}^2)_2&=& \beta_1{\vec \rho}^2({{\vec p}^2})^2+\beta_2 ({\vec \rho}{\vec p})^2{\vec p}^2.
\eea
The anomalous  deformed commutator, given by
\bea\label{anomcas}
\relax [{L_i}^{\cal F} , ({\vec L}^2)^\sharp]_{\cal F} &=& 4 \rho_i{\vec p}^2 +2i\epsilon_{kjl}\rho_kp_lL_j-2i({\vec \rho}{\vec p}) \epsilon_{ijl}p_lL_j +\nonumber\\&&
\alpha_1 i\epsilon_{ijk}\rho_jp_k({\vec p}{\vec L})+\alpha_2 i\epsilon_{ijk}({\vec p}^2)\rho_jL_k+
(2\beta_2-\alpha_2)i\epsilon_{ijk}\rho_jp_k({\vec \rho}{\vec p}){\vec p}^2,\nonumber\\
&&
\eea
does not depend on the $\beta_1$ coefficient. The minimal anomaly is recovered by setting
\bea
&\alpha_1=\alpha_2=\beta_2=0.&
\eea
The minimal anomaly is therefore given by the first line in the r.h.s. of (\ref{anomcas}).\par
The deformed rotational anomaly can be discussed for more general potential terms. Let us consider the addition of an
an anharmonic quartic term, given by
\bea\label{anham}
B= B_0 &=& \lambda \frac{({\vec x}^2)^2}{\hbar^3},
\eea
to the harmonic oscillator potential. In the above formula $\lambda$ is a positive coupling constant which, for simplicity, will be set equal to $1$. The $\frac{1}{\hbar^3}$ factor is introduced, as recalled at the end of Section {\bf 3}, in order to make (\ref{anham}) a primitive element of an infinite-dimensional Lie algebra $g$ containing the oscillator algebra $osc$ (\ref{ocomm}) as a finite subalgebra (since $\hbar$ is a central element of the Lie algebra $g$ one can evaluate it by setting, as usual, $\hbar=1$).\par
The $B^\sharp$ expansion starting from $B_0$ given in (\ref{anham}) produces a minimal anomaly for $B_k=0$ for $k\geq 5$.
By using the iterative procedure (\ref{iteranom}), we obtain that
\bea
B_1\equiv\left(\frac{({\vec x}^2)^2}{\hbar^3}\right)_1&=&\frac{ 4i}{\hbar^2}(p_i({\vec \rho}{\vec x})-({\vec \rho}{\vec x})x_i){\vec x}^2
\eea
gives the minimal anomaly of the first order, expressed by
\bea
(A_i)_1 &=& \frac{8}{\hbar}(2\rho_i{\vec x}^2-x_i({\vec \rho}{\vec x})).
\eea
At the next order we have that
\bea
B_2\equiv\left(\frac{({\vec x}^2)^2}{\hbar^3}\right)_2&=&\frac{ 2}{\hbar}\left(
2(\epsilon_{ijk}\rho_ip_jx_k)(\epsilon_{lmn}\rho_lp_mx_n)-({\vec \rho}{\vec p})^2{\vec x}^2\right)+
i\alpha ({\vec \rho}{\vec p})({\vec \rho}{\vec x})
\eea
produces an anomalous term $(A_i)_2$, given by
\bea\label{ambanom}
\left(A_i\right)_2 &=& \frac{16 i}{\hbar} \rho_i(\epsilon_{ljk}\rho_l p_jx_k) ({\vec p}{\vec x})-
16 \rho_i (\epsilon_{ljk}\rho_l p_j x_k)
+\nonumber\\
&&
-(4+\alpha )({\vec \rho}{\vec p})(\epsilon_{ijk} \rho_j x_k)
-(12+\alpha )(\epsilon_{ijk} \rho_j p_k)({\vec \rho}{\vec x})
.
\eea
The choice $\alpha=-4$ (respectively, $\alpha=-12$) makes disappear the third (fourth) term in the right hand side.\par
In the final example we consider the deformation of the Coulomb potential
\bea
B=B_0&=& \frac{1}{r},
\eea
with $r=\sqrt{{\vec x}^2}$.\footnote{It is worth to mention that it is consistent to produce the infinite-dimensional Lie algebra $g$ of primitive elements obtained by repeatedly applying the commutation relations to the generating elements $\hbar, x_i, p_i, \frac{1}{r}$.} \par
Unlike the expansion for the (\ref{anham}) anharmonic oscillator potential, the iterative procedure in this case never stops ($B_k\neq0$ at all orders).\par
Due to
\bea\label{tibo}
T_i(B_0)&=& i\hbar\rho_k(p_ix_k-p_kx_i)\frac{1}{r^3}+\hbar^2(\frac{\rho_i}{r^3}-3x_i\frac{{\vec\rho}{\vec x}}{r^5}),
\eea
we can set
\bea
B_1=\left(\frac{1}{r}\right)_1 &=& \alpha \hbar \epsilon_{ljk}\rho_lp_jx_k\frac{1}{r^3},
\eea
so that
\bea
L_i(B_1) &=& i \alpha\hbar (-{\vec \rho}{\vec p} x_i+p_i{\vec \rho}{\vec x})\frac{1}{r^3}.
\eea
By choosing
\bea
\alpha&=& -1,
\eea
the term proportional to $\hbar$ in (\ref{tibo}) can be reabsorbed. On the other hand, one can easily see that the anomalous term
$\hbar^2(\frac{\rho_i}{r^3}-3x_i\frac{{\vec\rho}{\vec x}}{r^5})$ cannot be reabsorbed by the contributions
coming from the higher order terms $\left(\frac{1}{r}\right)_k$ for $k>1$.\par
Summarizing, we get
\bea
\left(\frac{1}{r}\right)^\sharp &=& \frac{1}{r}-\hbar\epsilon_{ljk}\rho_lp_jx_k\frac{1}{r^3} + O(\hbar^2),
\eea
satisfying the anomalous twist-deformed commutator (with minimal anomaly)
\bea
\relax \left[{L_i}^{\cal F} , \left(\frac{1}{r}\right)^\sharp\right]_{\cal F} &=& \hbar^2(\frac{\rho_i}{r^3}-3x_i\frac{{\vec\rho}{\vec x}}{r^5})+O(\hbar^3).
\eea
\section{Conclusions}

In this work we investigated the Non-commutative Quantum Mechanics as a result of an abelian Drinfeld twist provided by formula (\ref{abeliantwist}). The twist deforms the Hopf algebra defined on the Universal Enveloping Algebra of a suitable Lie algebra. The Lie algebra under consideration, named {\it dynamical Lie algebra}, contains composite operators of the Heisenberg algebra operators $x_i, p_i, \hbar$. Nevertheless, these composite operators should be treated
as primitive elements, i.e. as generators, of the dynamical Lie algebra. Besides the Hamiltonian, the dynamical Lie algebra also includes, among its generators, the momenta $p_i$
in order to have a well-defined action of the (\ref{abeliantwist}) twist on its Enveloping Algebra endowed with the Hopf algebra structure. The dynamical Lie algebra
is either finite if its primitive elements result from composite operators at most quadratic in $x_i,p_i$; it is an infinite Lie algebra otherwise.  \par
We gave motivations for the use of ${\cal F}$-deformed generators and ${\cal F}$-deformed commutators in dealing with the twist-deformed Enveloping Algebra and pointed out the connection between twist and non-commutativity, which is provided by the equations (\ref{ncemer}) and (\ref{nctwist}), with $\theta_{ij}$ a constant operator belonging
to ${\cal U}(h_B(3))$. \par
The ${\cal F}$-deformed angular momenta ${L_i}^{\cal F}$ close the $so(3)$ algebra under ${\cal F}$-deformed brackets. On the the other hand, several operators which
at the undeformed level are rotationally invariant are anomalous in the deformed case, with the anomaly expressed by equation (\ref{anomaly}). We discussed various examples of anomalous operators. For the (deformed) $3D$ harmonic oscillator potential the anomaly is a linear constant operator, see (\ref{constanom}). In more general cases the anomaly is an operator which belongs to the Enveloping Algebra and is not necessarily constant. The concept of ``minimal anomaly" can be introduced. It corresponds to the specific choice, made so that to minimize the r.h.s. of equation (\ref{anomaly}), of the higher-order terms in the ${\vec \rho}$ Taylor-expansion of the deformed operator $B^\sharp$. In some cases the notion of minimal anomaly becomes ambiguous. This is the case for instance of formula (\ref{ambanom}). A (different) term contributing to the anomaly is eliminated by choosing the arbitrary parameter $\alpha$ to be either $\alpha=-4$ or $\alpha=-12$.\par
We also discussed the anomalous property of the twist-deformed Coulomb potential and of the twist-deformed $so(3)$ composite Casimir operator ${\vec L}^2$,
regraded as belonging to ${\cal U}(e(3))$, the Enveloping Algebra of the three-dimensional Euclidean algebra.\par
In a different (not involving the Drinfeld twist) context from ours, non-commutative Quantum Mechanics has been studied in several works, see e.g. \cite{cst},
where the non-commutative hydrogen atom was discussed, and \cite{glr}. In \cite{bg} (see also {\cite{der} and \cite{dh}) the classical counterpart of the non-commutative quantum mechanics is shown to be a constrained system. In \cite{gk} investigations of a dynamical (i.e. non-constant) non-commutative matrix
$\theta_{ij}$ were made.
\\ ~
\\ ~
\par {\large{\bf Acknowledgments}}{} ~\\{}~\par

We have profited of clarifying discussions with P. Aschieri on the properties of the ${\cal F}$-deformed commutators.
We acknowledge several useful discussions with P. G. Castro.
Z. K. and F. T. are grateful to the S. N. Bose National Center for Basic Sciences of Kolkata for hospitality.
B. C. acknowledges a TWAS-UNESCO associateship appointment at CBPF
and CNPq for financial support.
The work was supported by Edital Universal CNPq, Proc. 472903/2008-0 (Z.K., F.T).

~\\{}~\par

{{\Large \bf Appendix}}

~\par

~\par
The use of the ${\cal F}$-deformed brackets (\ref{Fbrackets}) in association with Non-commutative theories can be argued to appear naturally according to the following heuristic considerations. Let us consider a general Lie algebra ${\bf g}$ corresponding to a group
$G$ and its Universal Enveloping Lie algebra ${\cal U}({\bf g})$. Let the generators satisfy the commutation relations
\bea
[T_a,T_b]&=&if_{ab}^cT_c.
\eea
The adjoint representation of the group $G$ is obtained by the adjoint action of the group element,
according to
\bea\label{adjact}
T_a \rightarrow gT_ag^{-1}&=& D_{ab}T_b.
\eea
The matrices $D$ provide the adjoint representation for $g\in G$ and satisfy $D(g_1)D(g_2)=D(g_1g_2)$. In the infinitesimal version, near the identity, we can express $g$ as
\bea
g&\approx& {\bf 1}+i\omega_aT_a,
\eea
so that the above transformation (\ref{adjact}) reads as
\bea
T_a \rightarrow T'_a&=&T_a+\delta T_a,
\eea
with $\delta T_a=i\omega_b[T_b,T_a]$.

Before we generalize this construction to the NC case, we need to recast the commutative case itself in a
bit different setting. For that we can work with the group $G$, endowed with a Hopf group-algebra structure \cite{asc}. In the undeformed case, the undeformed
coproduct of $g$ is $\Delta_0(g)=g\otimes g$ and the antipode is $S(g)=g^{-1}$. Near the identity, it is equivalent to the following assignments of antipodes to the Lie algebra generators
$T_a$ and the identity ${\bf 1}$:
\bea
&S(T_a)=-T_a,\quad\quad S({\bf 1})={\bf 1}.
\eea
Now one can easily see that the finite form of the adjoint action on the group element is
\bea\label{adjactant}
T_a\rightarrow gT_ag^{-1}&=&gT_aS(g).
\eea
It is therefore associated with the coproduct $\Delta_0(g)=g\otimes g$.\par
When considering again $g$ to be close to the identity, the corresponding infinitesimal version of the coproduct is given by
\bea
\Delta_0(g)&=&{\bf 1}\otimes{\bf 1}+i\omega_b \Delta_0 (T_b),
\eea
where
\bea
\Delta_0 (T_b)&=& T_b\otimes {\bf 1}+{\bf 1}\otimes T_b
\eea
is the undeformed coproduct of the Lie-algebra generator $T_b$.
Just like ({\ref{adjactant}) the infinitesimal transformation of $T_a$ can be written with the help of the antipode
\bea
T_a \rightarrow T'_{a}&=&{\bf 1} T_a S({\bf 1})+i\omega_b(T_b T_a S({\bf 1})+{\bf 1} T_a S(T_b))= T_a+i\omega_b [T_b,T_a].
\eea
From this perspective the commutator brackets are associated with the undeformed coproduct.
Once this standpoint is adopted, it is natural to expect that the deformed coproduct arising in NC theories should be associated with the deformed
brackets.

To that end, let us consider the deformed coproduct obtained by the (\ref{abeliantwist}) Drinfeld abelian twist expressing the usual Moyal type of noncommutativity. We have
\bea
\Delta_0 (g)\rightarrow \Delta_{\cal F}(g) &=&{\cal F}\Delta_0(g) {\cal F}^{-1}
\eea
with ${\cal F}$ given by (\ref{abeliantwist}). Considering again a group element close to the identity, we
can write
\bea
\Delta_{\cal F}(g)&=& {\bf 1}\otimes{\bf 1}+ i\omega_b \Delta_{\cal F}(T_b).
\eea
Let the corresponding deformed coproduct for the Lie algebra generator $T_b$  be denoted in the Sweedler's notation (\ref{sweed}),
as
\bea
\Delta_{\cal F}(T_b)&=&\xi_1 \otimes \xi_2.
\eea
Repeating the same steps as before it can be easily seen that, in its infinitesimal form, the  transformation rule corresponding to the deformed coproduct of a generic element $A\in {\cal U({\bf G})}$ is
\bea
A\rightarrow A'&=&A+ \delta A,
\eea
with
\bea
\delta A&=&i\omega_b [T_b,A]_{\cal F}.
\eea
The deformed bracket reads as
\bea
[T_b, A]_{\cal F}&=& \xi_1 A S(\xi_2).
\eea
However, it can be easily seen that the algebra will not close under the deformed brackets, unless the original generators $T_a$'s are deformed further as \cite{asc} (see also formula (\ref{defgen}))
\bea
T_a^{\cal F}&=& {\overline f}^\alpha (T_a){\overline f}_\alpha.
\eea
These deformed generators span a linear subspace of the deformed Hopf algebra ${\cal U}^{\cal F}({\bf g})$ and their deformed brackets induce on them a Lie-algebraic structure. One has to note at least two important differences of these deformed generators and brackets in contrast to the undeformed ones.
Firstly, the exponentiation of the deformed generators does not yield elements of the Lie group, so that only the infinitesimal
version of the symmetry transformations are considered. Secondly, the deformed brackets are not manifestly antisymmetric, as it can be easily verified.

\end{document}